\documentclass[aip,jmp,showpacs,showkeys,amsmath,reprint,onecolumn]{revtex4-1}

\begin{document}

   \title{Centre-of-mass and internal symmetries in classical relativistic systems}

  \author{Uri Ben-Ya'acov}

 \affiliation{School of Engineering, Kinneret Academic College on
   the Sea of Galilee,\\D.N. Emek Ha'Yarden 15132, Israel}

   \email{uriby@tx.technion.ac.il}

    \date{\today}

\begin{abstract}
The internal symmetry of composite relativistic systems is discussed. It is demonstrated that Lorentz-Poincar\'{e} symmetry implies the existence of internal moments associated with the Lorentz boost, which are Laplace-Runge-Lenz (LRL) vectors. The LRL symmetry is thus found to be the internal symmetry universally associated with the global Lorentz transformations, in much the same way as internal spatial rotations are associated with global spatial rotations. Two applications are included, for an interacting 2-body system and for an interaction-free many-body system of particles. The issue of localizability of the relativistic CM coordinate is also discussed.
\end{abstract}

Submitted to J. Math. Phys.

\pacs{03.30.+p, 11.30.Cp, 11.30.Ly, 45.05.+x, 45.20.-d, 45.50.-j, 45.90.+t}

\keywords{Classical relativistic dynamics, Internal relativistic dynamics, relativistic internal symmetry, relativistic centre-of-mass, Runge-Lenz vector, Runge-Lenz symmetry}

\maketitle

\section{Introduction}

Separation of the internal dynamics of composite relativistic
systems from their global dynamics is one of the long-standing yet
not fully (or satisfactorily) solved issues in fundamental theoretical physics. Lorentz-Poincar\'{e} symmetry implies the constancy of the total linear momentum $P^\mu$ and total angular momentum $J^{\mu\nu}$, but not the uniqueness of the centre-of-mass (CM) coordinate. Thus, various definitions have been suggested for the relativistic CM, based on different requirements (see, \textit{e.g.}, Refs.~\onlinecite{Pryce,NewtonWigner49,Fleming65a} for early publications on the subject, or Refs.~\onlinecite{ALP02,Alba07} for recent publications with an extensive bibliography covering its history).

In non-relativistic systems, the separation of the total dynamics into CM motion and internal dynamics is incorporated just in
splitting the total energy and the total angular momentum into
sums of CM-terms and internal terms. In relativistic systems we
know, correspondingly, to isolate the internal energy as the total
invariant mass of the system, and to extract from $J^{\mu\nu}$ its
part which is responsible for internal rotations -- the spatial internal angular momentum tensor (see Eq.~(\ref{eq: ell mu nu def}) below).

However, internal symmetry is more than just rotations. It is well known that in 2-body Newtonian Kepler-Coulomb systems there exists the Laplace-Runge-Lenz (LRL) vector as a constant of the motion. Knowledge of the
LRL vector amounts to having a full solution for the
configuration of the system (details of orbit, etc.) \cite{Goldstein02} ; in the
corresponding quantum systems the LRL vector provides a
very elegant means for obtaining the full quantum picture of the
system (as in the case of the hydrogen atom) \cite{McIntosh71}. Although still regarded by many as corresponding to 'accidental' or 'hidden' symmetry, particular only to $1/r$ potentials, it is known already for a long time that LRL vectors exist in all rotationally symmetric systems \cite{BacryRS66,Fradkin67,Mukunda67a,Peres79,HolasMarch90}, including various relativistic models as well
\cite{Fradkin67,Yoshida88b,ArgSanz84,DVNur90,Horwitz93,Duviryak96,JohnsonLipp50,Biedenharn62,KhachidzeKhelash06,KhachidzeKhelash08}. As in classical Kepler-Coulomb systems, these LRL vectors may be constructed to be constant\cite{LRLgeneral}, and they generate, together with the internal angular momentum, $SO(4)$ or $SO(3,1)$ symmetry according to the energetical state of the system. It follows, therefore, that the mere existence of the LRL vectors is a direct and essential consequence of the existence of internal rotational symmetry, and the $SO(4)$ or $SO(3,1)$ symmetry thus generated is a natural extension of the rotational symmetry\cite{BacryRS66,Fradkin67,Mukunda67a,LRLgeneral}.

Considering composite relativistic systems, since Lorentz-Poincar\'{e} symmetry implies internal rotational symmetry, the foregoing argument then leads to expect the appearance of LRL vectors in all relativistic systems. In fact, not only do LRL vectors exist in relativistic systems (and therefore participate in generating the internal symmetry there), their very origin lies within the relativistic domain : A number of years ago it was discovered by Dahl \cite{Dahl97,Dahl68} that the Newtonian LRL vector emerges in the computation of the Lorentz boost in the post-Newtonian approximation of electromagnetic or gravitational 2-body systems. Although it is only the Newtonian LRL vector this an essentially relativistic result, because it appears in terms of order $1/c^2$, vanishing in the full non-relativistic limit when the Lorentz boost becomes the Galilei boost. Dahl's results were recently re-established and extended to other systems, including fully relativistic non-interacting 2-body systems \cite{LocCM} and arbitrary post-Newtonian centrally symmetric 2-body systems \cite{LRLgeneral}.

Combining these two characteristics -- the existence of LRL vectors for all rotationally symmetric systems together with their relativistic origin -- it is then natural to expect that LRL vectors form an integral part of the internal relativistic symmetry. Also, since the spatial part of the relativistic CM coordinate is derived from the Lorentz boost, from which also the LRL vector is derived, it is natural to expect that
the LRL vector plays a significant r\^{o}le in the determination of the relativistic CM. These subjects which so far were regarded as completely distinct -- determination
of the relativistic centre-of-mass, relativistic internal symmetry and LRL vectors -- appear therefore to form part of one and the same story.

So far, LRL vectors and the associated symmetry were considered, even in systems with relativistic features, as extensions of Newtonian systems rather than from the stand-point of full Lorentz-Poincar\'{e} symmetry. The purpose of the present paper is thus to discuss in detail the internal symmetry in composite Lorentz-Poincar\'{e} symmetric systems, in a manifestly covariant manner in Minkowski space-time, tracing the appearance of LRL vectors in these systems, all in association with the issue of determining the CM coordinate and its properties. The paper extends and completes recent publications on the subject\cite{LRLgeneral,LocCM}.

The paper is composed in the following of three main parts. First, in Secs.~\ref{sec: gen CM}, \ref{sec: rel LRL} and \ref{sec: rel int} are discussed in generic terms the appearance of the LRL vectors and their association with the CM coordinate. It is shown that the mere existence of Lorentz-Poincar\'{e} symmetry implies the existence of LRL vectors, which together with the internal angular momentum generate the internal symmetry of the system, at least for 2-body systems. The spatial component of the CM coordinate contains an internal part which is proportional to the LRL vector of the system, and the type of the symmetry depends only on the energetical state of the system, being $SO(4)$ or $SO(3,1)$ for bound or unbound systems, respectively.

The methods developed in the first part are then applied in Secs.~\ref{sec: SVint} and \ref{sec: non-int many} to two particular systems. As an illustration of the application of these methods, a 2-body system with light-like antisymmetric scalar-vector interaction is considered in Sec.~\ref{sec: SVint}. This is a relatively simple model, so chosen to keep the exposition simple and clear. Then, since the procedure developed in the first part allows the definition of many-body LRL vectors, it is applied in Sec.~\ref{sec: non-int many} to non-interacting many-body systems, thereby offering a first step into the generalization of the LRL symmetry also to interacting many-body systems.

Finally. the implications of the foregoing results for the relativistic CM coordinate, in particular the issue of its localizability and its association with properties of the LRL vector of the system, are discussed in section~\ref{sec: CMloc}. Final discussion and concluding remarks are given in Sec.~\ref{sec: concrem}.

\textit{Notation}. In the following we consider dynamics
described in a Minkowski space-time $\{x^\mu\}$, $\mu=0,1,2,3$
with metric tensor $g_{\mu\nu}=\text{diag}(-1,1,1,1)$. Spatial (3D)
vectors are denoted by an arrow. Dot product is $A \cdot B = A^\mu B_\mu$. The unit fully anti-symmetric (Levi-Civita) pseudo-tensor is
$\varepsilon^{\mu\nu\lambda\rho} = -\varepsilon_{\mu\nu\lambda\rho} = 1$ for $(\mu\nu\lambda\rho)$ an even permutation of $(0,1,2,3)$. It is also assumed throughout
that $c=1$ unless specified otherwise. An orthogonality subscript ($\bot$) implies the component of a 4-vector perpendicular to the total linear momentum $P^\mu$.

\section{\label{sec: gen CM}The CM coordinate and decomposition of the total angular momentum}

We start by discussing the construction of the CM coordinate of general classical (non-quantum) composite relativistic systems and the decomposition of the total angular momentum into CM-terms and internal terms. We consider systems of $N$ point particles, with masses $m_a$ and moving on the trajectories $x^\mu = x_a^\mu(\tau_a)$, $a=1,...,N$, $\tau_a$ being the proper time of the $a$-th particle.

With constant total linear momentum, the space-time trajectory of the CM coordinate is expected to be a straight line in the direction of $P^\mu$, and it may always be written as the centroid
 \begin{equation} \label{eq: X trajectory}
 X^\mu(\tau) = X_o^\mu + \tau \cdot \frac{P^\mu}{M}
 \end{equation}
where $M = \sqrt{- P^\mu P_\mu}$ is the invariant mass of the system, $\tau$ is the CM
proper-time, both Lorentz scalars, and $X_o^\mu$ is a constant 4-vector, identified as the spatial CM coordinate. Appropriately fixing the zero of
$\tau$, $X_o^\mu$ may be assumed orthogonal to $P^\mu$ without
loss of generality, $ X_o \cdot P = 0$. The properties of $\tau$
as an observable were discussed in Refs.~\onlinecite{Internaltime,Inttimedil}; those of $X_o$ are discussed in the present article.

With Lorentz-Poincar\'{e} symmetry, the common view maintains that the CM coordinate and the internal symmetry of relativistic systems should be deduced from $P^\mu$
and $J^{\mu\nu}$ alone, together with the particles' spins, if the latter exist\cite{Pryce,Fleming65a,ALP02,Alba07}. For manifestly Lorentz-covariant (observer independent) expressions for the CM coordinate, the colloquial choice so far has always been the so called \textit{centre-of inertia}
 \begin{equation} \label{eq: XI def}
 X_\text{I}^\mu \equiv - \frac{J^{\mu\nu} P_\nu}{M^2}
 \end{equation}
This choice was backed up by the fact that $X_\text{I}^\mu$ is the unique solution possible for $X_o^\mu$ if it is required to be formed of $P^\mu$ and $J^{\mu\nu}$ alone \cite{Fleming65a,ALP02}. It ceases, however, to be the solution when dependence on internal observables is taken into account, as is manifestly discussed in the following.

Once the CM coordinate (Eq.~(\ref{eq: X trajectory})) is assumed to be known, the total angular momentum may always be split into combination of orbital (CM) and internal parts,
 \begin{equation} \label{eq: split J}
J^{\mu\nu} = X^\mu P^\nu - X^\nu P^\mu + j^{\mu\nu} = X_o^\mu P^\nu
- X_o^\nu P^\mu + j^{\mu\nu}
 \end{equation}
$j^{\mu\nu}$ is the internal angular momentum, relative to the centre-of-mass. From Eq.~(\ref{eq: split J}) it follows that it is a constant. Out of the 6 components of $j^{\mu\nu}$ 3 are independent of the CM-coordinate, fully determined by $J^{\mu\nu}$ and $P^\mu$ via the condition
 \begin{equation} \label{eq: J-j cond}
\epsilon_{\mu\nu\lambda\rho} \left(J^{\mu\nu} - j^{\mu\nu} \right)
P^\lambda = 0
 \end{equation}
which follows from Eq.~(\ref{eq: split J}). These components of
$j^{\mu\nu}$ constitute the spatial internal angular
momentum tensor
 \begin{equation} \label{eq: ell mu nu def}
\ell^{\mu\nu} \equiv \Delta^\mu_\lambda \Delta^\nu_\rho
J^{\lambda\rho} = - \left( {J^{\mu\nu} P^\lambda + J^{\lambda\mu}
P^\nu + J^{\nu\lambda} P^\mu} \right) \frac{P_\lambda}{M^2} =
\Delta^\mu_\lambda \Delta^\nu_\rho j^{\lambda\rho}
 \end{equation}
where
 \begin{equation} \label{eq: gamma def}
\Delta^{\mu\nu} \equiv g^{\mu\nu} + \frac{P^\mu P^\nu}{M^2}
 \end{equation}
is the 3-D metric tensor in the spatial part of the CM reference
frame. $\ell^{\mu\nu}$ is certainly non-zero in the general case.

The remaining 3 components of $j^{\mu\nu}$ determine $X_o^\mu$. It
is convenient to define the vector
 \begin{equation} \label{eq: Q from j}
 Q^\mu \equiv \frac{j^{\mu\nu} P_\nu}{M^2}
 \end{equation}
which incorporates these components ($Q \cdot P = 0$, so $Q^\mu$ contains only 3 degrees of freedom). Inverting Eq.~(\ref{eq: split J}), $X_o^\mu$ is uniquely defined in terms of $P^\mu$, $J^{\mu\nu}$ and $Q^\mu$ :
 \begin{equation} \label{eq: Xo from J&j}
X_o^\mu = - \frac{\left(J^{\mu\nu} - j^{\mu\nu}\right) P_\nu}{M^2} = X_\text{I}^\mu + Q^\mu
 \end{equation}

In the following, we refer as \textit{internal} to observables that : 1) are invariant
under uniform translations ; 2) if not scalars, all their components are confined to the spatial part of the CM reference frame (the hyperplane perpendicular to $P^\mu$). Thus, all internal vectors ${\mathcal A}^\mu$ satisfy ${\mathcal A} \cdot P = 0$, with a corresponding relation for internal tensors.

$Q^\mu$ and $\ell^{\mu\nu}$ are then constant internal quantities. By Eq.~(\ref{eq: Xo from J&j}), $Q^\mu$ takes the r\^{o}le of a shift or displacement vector, telling us by how much the spatial part of the CM coordinate is removed from the centre-of-inertia. It is noted that $Q^\mu$ cannot be formed out of the Lorentz-Poincar\'{e} global generators $P^\mu$ and $J^{\mu\nu}$ alone, because the only internal quantity that they
can form is $\ell^{\mu\nu}$. A non-zero $Q^\mu$ thus challenges the common view that the relativistic CM should be constructed of $P^\mu$ and $J^{\mu\nu}$ alone \cite{Fleming65a,ALP02}. $Q^\mu$, if non-zero, is therefore a \textit{new} constant internal vector.

Expressed in terms of $\ell^{\mu\nu}$ and $Q^\mu$, $j^{\mu\nu}$ becomes
 \begin{equation} \label{eq: j decomposion}
 j^{\mu \nu} = \ell^{\mu\nu} - Q^\mu P^\nu + Q^\nu P^\mu
 \end{equation}
While $\ell^{\mu\nu}$ is responsible for (spatial) rotations relative to the CM reference frame, the remaining part of $j^{\mu\nu}$ may be regarded as the internal moment corresponding to the Lorentz boost. In the following we demonstrate that $Q^\mu$ does indeed exist and is non-zero in the general case, being proportional to the LRL vector of the system, and is therefore responsible (together with $\ell^{\mu\nu}$) for internal symmetries. Combined together via Eq.~(\ref{eq: j decomposion}), $j^{\mu\nu}$ is then responsible for the fundamental internal symmetry of the system.

\section{\label{sec: rel LRL}Relativistic Internal (Laplace-Runge-Lenz) symmetry}

In the present section we discuss the internal symmetry induced by the global Lorentz-Poincar\'{e} symmetry. It has already been shown in the past \cite{BacryRS66,Fradkin67,Mukunda67a,LRLgeneral} that LRL vectors exist in general rotationally symmetric Newtonian-like systems, which generate together with internal rotations the internal symmetry. The picture will now be completed by showing that the mere existence of the global Lorentz-Poincar\'{e} symmetry implies the existence of LRL vectors, which, in the same way, generate the internal symmetry together with internal rotations.

The basis for the generalization of the LRL symmetry to general rotationally symmetric system has been shown \cite{LRLgeneral} to be incorporated in two basic propositions. Due to their importance, let us rephrase and prove these propositions in manifestly Lorentz-covariant terms.

Even in the absence of clear and unique definition of canonical
phase-space variables in classical relativistic systems, still the behaviour of
various observables under global transformations of Minkowski space-time determines certain relations that any plausible definition of Poisson
brackets (PB) must satisfy. These alone suffice to determine the generic properties of the internal Lie-Poisson algebra. Let $\{{\mathcal A},{\mathcal B}\}$ denote the PB of any two
observables ${\mathcal A}$ and ${\mathcal B}$. If $\delta {\mathcal G}$ is the
generator of an infinitesimal space-time transformation, and $\delta {\mathcal
A}$ is the variation of an observable ${\mathcal A}$ under that
transformation, then the PB should satisfy the relation
 \begin{equation} \label{eq: PB inf trans}
 \delta {\mathcal A} = \{{\mathcal A},\delta {\mathcal G}\} \, ,
 \end{equation}
together with the standard rules of Lie-Poisson algebras
\cite{MarsdenRatiu99}, namely :
\begin{subequations}\label{eq: PB rules}
 \begin{eqnarray}
&& \text{Antisymmetry \, :} \hskip 90pt \{ {\mathcal A},{\mathcal B}\} = - \{ {\mathcal B},{\mathcal A}\}
 \label{eq: AS PB} \\
&& \text{Jacoby identity \, :} \hskip 80pt \{ {\mathcal A},\{ {\mathcal B},{\mathcal C}\} \} + \{ {\mathcal B},\{ {\mathcal
C},{\mathcal A}\} \} + \{ {\mathcal C},\{ {\mathcal A},{\mathcal B}\} \} = 0
 \label{eq: PB Jac} \\
&& \text{Product ("Leibnitz") rule \, :} \hskip 20pt \{{\mathcal A},{\mathcal B}{\mathcal C}\} = \{{\mathcal A},{\mathcal B}\} {\mathcal C} + \{{\mathcal A},{\mathcal C}\}{\mathcal B}
 \label{eq: PB product}
 \end{eqnarray}
\end{subequations}

In the following the fundamental PB are those of the Lorentz-Poincar\'{e}
Lie-Poisson algebra,
 \begin{eqnarray} \label{eq: PB PJ}
&& \{P^\mu,P^\nu\} = 0 \quad , \qquad \{J^{\mu\nu},P^\lambda\} =
g^{\mu\lambda} P^\nu -
g^{\nu\lambda} P^\mu  \nonumber \\
&&  \{J^{\mu\nu},J^{\lambda\rho}\} = g^{\mu\lambda} J^{\nu\rho} -
g^{\nu\lambda} J^{\mu\rho} - g^{\mu\rho} J^{\nu\lambda} +
g^{\nu\rho} J^{\mu\lambda}
  \end{eqnarray}
Besides these relations, no specific canonical structure is
assumed. The PB of any other quantity which is constructed from
$P^\mu$ and $J^{\mu\nu}$ are easily computed using the
product rule (\ref{eq: PB product}) and the derivative rule
 \begin{equation} \label{eq: derivative rule}
  \left\{{\mathcal A},f({\mathcal B})\right\} = \left\{{\mathcal A},{\mathcal B}\right\} f'({\mathcal B})
 \end{equation}
which follows from it. The PB of other observables are deduced from their transformation
properties. In particular, all 4-vectors $V^\mu$ satisfy
 \begin{equation} \label{eq: PB rot V}
\left\{ V^\mu,J^{\nu\lambda} \right\} = g^{\mu\lambda}
 V^\nu - g^{\mu\nu} V^\lambda
 \end{equation}
and all internal observables $\mathcal A$ satisfy $\left\{ {\mathcal A} ,P^\mu \right\} = 0$.

The spatial internal angular momentum tensor $\ell^{\mu\nu}$ is defined from $P^\mu$ and $J^{\mu\nu}$ via Eq.~(\ref{eq: ell mu nu def}). Dual to it is the vector (which is proportional to the well-known Pauli-Lubanski vector)
 \begin{equation} \label{eq: ell mu def}
\ell^\mu \equiv \frac{1}{2}\epsilon^{\mu\nu\lambda\rho} J_{\nu\lambda}
U_\rho = \frac{1}{2}\epsilon^{\mu\nu\lambda\rho} \ell_{\nu\lambda}
U_\rho
 \end{equation}
where $U^\mu = P^\mu/M$ is the unit 4-velocity vector of the
CM-frame, with the inverse duality relation
 \begin{equation} \label{eq: dual ell}
\ell^{\mu\nu} = \varepsilon^{\mu\nu\lambda\rho} \ell_\lambda
U_\rho
 \end{equation}
$\ell^\mu$ is also an internal quantity. The self PB of $\ell^{\mu\nu}$ are
\begin{subequations}\label{eq: PB ell}
 \begin{eqnarray}
&& \left\{ \ell^{\mu\nu},\ell^{\lambda\rho} \right\} =
\Delta^{\mu\rho} \ell^{\lambda\nu} - \Delta^{\nu\lambda}
\ell^{\mu\rho} + \Delta^{\nu\rho} \ell^{\mu\lambda} -
\Delta^{\mu\lambda} \ell^{\rho\nu}
 \label{eq: PB ell munu} \\
 \text{with the corresponding ones of $\ell^\mu$} &&
 \nonumber \\
&& \left\{ \ell^\mu ,\ell^\nu \right\} = \ell^{\mu\nu}
 \label{eq: PB ell mu}
 \end{eqnarray}
\end{subequations}
From the rotational PB (\ref{eq: PB rot V}), it follows that the PB of any internal vector ${\mathcal A}^\mu$ with $\ell^{\mu\nu}$ are
\begin{subequations}\label{eq: A ell PB}
 \begin{eqnarray}
&& \left\{ {\mathcal A}^\mu,\ell^{\nu\lambda} \right\} = \Delta^{\mu\lambda}
{\mathcal A}^\nu - \Delta^{\mu\nu} {\mathcal A}^\lambda
 \label{eq: A ell munu PB} \\
 \text{for $\ell$ in tensor form, or} &&
 \nonumber \\
&& \left\{ \ell^\mu ,{\mathcal A}^\nu \right\} = \left\{ {\mathcal A}^\mu ,\ell^\nu
\right\} = \varepsilon^{\mu\nu\lambda\rho} {\mathcal A}_\lambda U_\rho
 \label{eq: A ell mu PB}
 \end{eqnarray}
\end{subequations}
for $\ell$ in vector form, indicative of the fact that $\ell^{\mu\nu}$ is the generator of internal rotations in the CM reference
frame. It is clear that the PB (\ref{eq: PB ell}) and
(\ref{eq: A ell PB}) maintain the property of being internal.

Let $K^\mu$ be an internal vector observable. $K^2 = K^\mu K_\mu$ and $K \cdot \ell = K^\mu \ell_\mu$, being internal scalar observables, certainly satisfy  $\left\{ \ell^\mu,K^2 \right\} = 0$ and $\left\{ \ell^\mu, K \cdot \ell \right\} = 0$. Then it may be shown that :

\vskip10pt \noindent \textbf{proposition 1} \,
\textit{The PB $\left\{ K^\mu,K^\nu \right\}$ are proportional to $\ell^{\mu\nu}$ if and only if $\left\{ K^\mu,\ell \cdot K \right\} = 0$.}

\vskip10pt \noindent \textbf{Proof} \, $K^\mu$, being an internal vector, satisfies $K \cdot P = 0$ and $\left\{K^\mu,P^\nu\right\} = 0$. Then, by the basic PB rules (\ref{eq: PB rules}) it follows that the self PB $\left\{ K^\mu,K^\nu \right\}$ constitute an internal tensor. Being anti-symmetric, these PB may always be written as $\left\{ K^\mu,K^\nu \right\} = \varepsilon^{\mu\nu\lambda\rho} \Lambda_\lambda U_\rho$ with $\Lambda^\mu$ some internal vector. Therefore, applying Eq.~(\ref{eq: A ell mu PB}) for $K^\mu$, it follows that
 \begin{equation} \label{eq: PB K-ellK}
\left\{ K^\mu,\ell \cdot K \right\} = \left\{ K^\mu ,\ell^\nu
\right\} K_\nu + \left\{ K^\mu,K^\nu \right\} \ell_\nu = \left\{
K^\mu,K^\nu \right\} \ell_\nu = \varepsilon^{\mu\nu\lambda\rho} \ell_\nu \Lambda_\lambda U_\rho
 \end{equation}
Since both $\ell^\mu$ and $\Lambda^\mu$ are internal vectors, it follows from Eq.~(\ref{eq: PB K-ellK}) that $\left\{ K^\mu,\ell \cdot K \right\}$ vanishes iff $\ell^\mu$ and $\Lambda^\mu$ are parallel. Thus, using the duality relation (\ref{eq: dual ell}), follows the proposition. \hskip220pt QED

\vskip10pt

In 2-body systems any constant internal scalar observable must be a function only of the total mass $M$ and of $\ell^2 \equiv \ell^\mu \ell_\mu = \ell^{\mu\nu} \ell_{\mu\nu} /2$. Let $K^\mu$ be an internal vector with constant squared magnitude $K^2$. Then necessarily $K^2$ must be a function of $M$ and $\ell^2$, say $K^2 = F\left(M,\ell^2\right)$. For more general systems, constant internal scalar observables need not be functionally dependent on $M$ and $\ell^2$ only. Still we may consider those vectors $K^\mu$ for which the product $\ell\cdot K$ is $ K$-invariant, $\{K^\mu,\ell\cdot K \} = 0$, and their magnitude $K^2$ is some function $K^2 = F\left(M, \ell^2, {\mathcal A}\right)$ where ${\mathcal A}$ stands for any internal scalar observable which is $K$-invariant, in the sense that  $\left\{ K^\mu,{\mathcal A} \right\} = 0$. Then we have

\vskip1pt \noindent \textbf{ proposition 2} \,
\textit{Let $K^\mu$ be an internal vector such that :
\begin{enumerate}
 \item
The product $\ell\cdot K$ is $K$-invariant
 \item
$K^2 = F(M,\ell^2,\mathcal{A})$
\end{enumerate}
Then the self PB of $K^\mu$ satisfy}
 \begin{equation} \label{eq: PB KK}
\left\{ K^\mu,K^\nu \right\} = - \frac{\partial\left[ K^2(\ell^2)
\right]}{\partial(\ell^2)} \ell^{\mu\nu}
 \end{equation}

\noindent \textbf{Proof} \,
From Eq.~(\ref{eq: A ell mu PB}) it also follows that
\[
\left\{ K^\mu ,\ell^2 \right\} = 2\left\{ K^\mu ,\ell^\nu
\right\}\ell_\nu = 2\varepsilon^{\mu\nu\lambda\rho} \ell_\nu
K_\lambda U_\rho = -2\ell^{\mu\nu}K_\nu
 \]
Let $\left\{ K^\mu,K^\nu \right\} = \alpha \cdot \ell^{\mu\nu}$.
Then
\[
\left\{ K^\mu ,K^2 \right\} = 2\left\{ K^\mu ,K^\nu \right\}K_\nu
= 2\alpha \cdot \ell^{\mu\nu} K_\nu
 \]
Combining the last two relations thus yields
\[
\left\{ K^\mu ,K^2 \right\} + \alpha \cdot \left\{ K^\mu ,\ell^2
\right\} = 0
 \]
from which follows, using the derivative rule (\ref{eq: derivative
rule}), Eq.~(\ref{eq: PB KK}). \hskip220pt QED

\vskip10pt

Any vector that satisfies the conditions of these propositions may be regarded as a \textit{relativistic LRL vector}. In 2-body systems there are 6 internal degrees of
freedom. 4 of them are contained in the total relativistic mass $M$ and the
internal angular momentum $\ell^{\mu\nu}$. The other 2 must be contained in a LRL vector, because it is always possible to construct a constant vector in the plane of motion \cite{LRLgeneral} (the plane defined by $\ell^{\mu\nu}$), and any constant scalar can only be a function of $M$ and $\ell^2$.

It follows, therefore, that the shift vector $Q^\mu$ must -- necessarily -- be a LRL vector which generates, together with $\ell^{\mu\nu}$, the internal symmetry of the system. This internal symmetry is governed by the PB (\ref{eq: PB ell}), (\ref{eq: A ell PB}) and (\ref{eq: PB KK}). With $j^{\mu\nu}$ being expressed in terms of $\ell^{\mu\nu}$ and $Q^\mu$ via Eq.~(\ref{eq: j decomposion}) it follows that $j^{\mu\nu}$ is the responsible for the fundamental internal symmetry of the system.

As is discussed in detail in Ref.~\onlinecite{LRLgeneral}, the nature of the symmetry is determined by Eq.~(\ref{eq: PB KK}). For any particular system the value of
 \begin{equation} \label{eq: eta}
\eta \equiv - \text{sign}\left[ \frac{\partial\left(K^2\right)}{\partial(\ell^2)} \right]
 \end{equation}
is the same for all LRL vectors, depending on the energetic state of the system : $\eta = +1$ or $-1$ for bound or unbound systems, respectively. The internal symmetry generated by $j^{\mu\nu}$ is then, respectively, $SO(4)$ or $SO(3,1)$, for bound or unbound systems. The transformations generated by the LRL vector change, for a given value of the total energy,
the internal angular momentum, thus changing the internal
configuration of the system -- how the particles move relative to
the centre-of-mass -- taking the system
from one orbit to another, with the same energy. Explicit knowledge of the LRL vector may be used, as it does for classical (Newtonian) systems,
to provide a full solution for the configuration of the system.

Finally, we have shown that $Q^\mu$ is a LRL vector for 2-body systems. The process of CM integration may be performed with any number of particles, and it is indeed shown in the following (section \ref{sec: non-int many}) that $Q^\mu$ is a LRL vector for free many-body systems. The same has already been shown for post-Newtonian many-body systems \cite{manybodyPN}. We may then conjecture that $Q^\mu$ is a LRL vector for arbitrarily large, composite relativistic systems with arbitrary internal interactions.

\section{\label{sec: rel int}Relativistic integration of the centre-of-mass}

In the present Section we bring together the results of Secs.~\ref{sec: gen CM} and \ref{sec: rel LRL}, providing an explicit procedure for the computation of the shift vector $Q^\mu$ and pointing at its relation with the LRL vector.

In principle, the single particle trajectories may be
parameterized each by a different time-like parameter, but for a common
evolution picture a common parameter is
required. Let $\sigma$ be such a common evolution parameter. Then
the single-particle trajectories are $x^\mu = x_a^\mu(\sigma)$.
Derivatives relative to $\sigma$ are denoted in the following by
an overdot, so that the particles' generalized velocities are
$\dot{x}^\mu = d{x^\mu}/d\sigma$.

Since $X_o^\mu  = X^\nu \Delta_\nu^\mu$ is constant, its
determining equation may be put in the form
 \begin{equation} \label{eq: dX/dsigma Del}
 \frac{dX^\nu}{d\sigma} \Delta_\nu^\mu = 0
 \end{equation}
The Lorentz-covariant generalization of the Newtonian CM, the 4-vector
 \begin{equation} \label{eq: XN def}
 X_\text{N}^\mu \equiv \frac{\sum_a {m_a x^\mu_a}}{M_o}
 \end{equation}
(with $M_o = \sum_a {m_a}$ as the Newtonian total mass) does not
satisfy Eq.~(\ref{eq: dX/dsigma Del}); however, its corresponding
derivative,
 \begin{equation} \label{eq: dXN/dsigma Del}
 \frac{dX_\text{N}^\nu}{d\sigma} \Delta_\nu^\mu = \frac{\sum_a m_a v_a^\mu}{M_o} \neq 0 \, ,
 \end{equation}
($v_a^\mu \equiv \dot{x}^\nu \Delta_\nu^\mu$ is the \textit{a-th}
particle's spatial velocity relative to the CM frame) vanishes in
the non-relativistic limit ($v_a/c \to 0$). Therefore, the
time-varying part of $X_\text{N}^\nu \Delta_\nu^\mu$ is purely
relativistic. In fact, in the non-relativistic limit $X_\text{N}^\nu \Delta_\nu^\mu - X_\text{I}^\mu \rightarrow 0$. Moreover,
whatever the vector $X_o^\mu$ may be, its behaviour under uniform
translations $x^\mu \rightarrow x^\mu + a^\mu$ must always be
 \begin{equation} \label{eq: unitrans Xo}
 X_o^\mu \rightarrow X_o^\mu + a^\mu + \frac{P \cdot a}{M^2} P^\mu \, ,
 \end{equation}
exactly like that of $X_\text{N}^\nu \Delta_\nu^\mu$, so the
difference $X_\text{N}^\nu \Delta_\nu^\mu - X_o^\mu$ is an internal
vector. Therefore, in order to determine $X_o^\mu$, we look for an
internal 4-vector $R^\mu$ which satisfies
 \begin{equation} \label{eq: R eqn}
 \frac{d R^\mu}{d\sigma} = \frac{dX_\text{N}^\nu}{d\sigma} \Delta_\nu^\mu
 \end{equation}
and vanishes in the non-relativistic limit, so that $X_o^\mu$ is given by
 \begin{equation} \label{eq: X iden}
 X_o^\mu = X_\text{N}^\nu \Delta_\nu^\mu - R^\mu
 \end{equation}

The centre-of-inertia (\ref{eq: XI def}), being a solution of
Eq.~(\ref{eq: dX/dsigma Del}), provides, via Eq.~(\ref{eq: X iden}), an
immediate solution of Eq.~(\ref{eq: R eqn}) in the form
 \begin{equation} \label{eq: R1 def}
 R_1^\mu = X_\text{N}^\nu \Delta_\nu^\mu - X_\text{I}^\mu
 \end{equation}
If $R_1^\mu$ was the only possible solution to Eq.~(\ref{eq: R eqn}),
then $X_o^\mu$ must be equal to $X_\text{I}^\mu$ and $Q^\mu$ must vanish. This, however, is not the case. Non-trivial solutions, independent of $R_1^\mu$, are possible, with corresponding non-zero shift vector $Q^\mu$ :
 \begin{equation} \label{eq: Q R1 R}
 Q^\mu = R_1^\mu - R^\mu
 \end{equation}

The process of solving Eq.~(\ref{eq: R eqn}) for the non-trivial solution, identifying the
shift vector (\ref{eq: Q R1 R}) and constructing consequently the
spatial CM component $X_o^\mu$ via Eq.~(\ref{eq: Xo from J&j}), is referred to in the
following as \textit{integration of the relativistic centre-of-mass}. The main property of the process, namely that the  time-varying part of the Newtonian CM for relativistic
systems is purely relativistic, vanishing in the limit $c \rightarrow \infty$, was first used by Dahl \cite{Dahl97} to compute $\vec X_\text{N}$ in the post-Newtonian
approximation of a 2-particle system interacting
electromagnetically or gravitationally. He showed, via integration
of an  equation like Eq.~(\ref{eq: R eqn}), that $\vec X_\text{N}$
results in a time-varying vector which is of order $1/c^2$ plus a
constant of integration $\vec X_o$. The surprising result was that
the constant $\vec X_o$ was not equal to $\vec X_\text{I}$, the
centre-of-inertia, but rather the difference $\vec X_o - \vec
X_\text{I}$ (now identified as the shift vector $\vec Q$) was found to be proportional to the LRL vector of the corresponding Newtonian system.

As a brief illustration of this procedure, let us recall, in an adapted way, Dahl's computation. Consider a 2-particle system with masses $m_1,m_2$, possible electrical charges $e_1,e_2$, described in the CM reference frame with post-Newtonian
EM/gravitational interaction \cite{LLFields75}. With $\vec r$ the
relative coordinate, $\vec v = \dot{\vec r}$ the relative velocity
and $\mu$ the Newtonian reduced mass, then in the post-Newtonian
approximation Eq.~(\ref{eq: R eqn}) becomes
 \begin{equation} \label{eq: dR/dt PN}
\frac{d\vec R}{dt} = \frac{\left( m_1 - m_2 \right)}{2M_o^2 c^2} \left[ \left( \mu v^2 + \frac{\kappa}{r} \right)\vec v + \frac{\kappa \left( \vec v \cdot \vec r \right)}{r^3} \vec r \right]
 \end{equation}
($\kappa = e_1 e_2$ or $\kappa = -G m_1 m_2$ for the electrical or
gravitational case, respectively; in the CM frame $R^o = 0$ hence
only the spatial part is relevant). Using the Newtonian equations
of motion (which are sufficient since the required overall
accuracy is $O(1/c^2)$)
\begin{equation} \label{eq: New eqs}
\mu \frac{d \vec v}{dt} = \frac{\kappa \vec r}{r^3}
\end{equation}
the square brackets in the rhs of Eq.~(\ref{eq: dR/dt PN}) may be
expressed as a total time derivative in either of two ways
\begin{equation} \label{eq: dR/dt PN eqn 1}
 \left( \mu v^2 + \frac{\kappa}{r} \right)\vec v + \frac{\kappa \left( \vec v \cdot \vec r \right)}{r^3} \vec r = \frac{d}{dt} \left[ \mu \left( \vec v
\cdot \vec r \right)\vec v \right] = \frac{d}{dt} \left[ \left( \mu v^2 + \frac{\kappa}{r} \right) \vec r \right]
\end{equation}
The post-Newtonian trivial solution, associated with the
centre-of-inertia via Eq.~(\ref{eq: R1 def}), is
 \begin{equation} \label{eq: R1 PN}
 \vec R_1 = \vec X_\text{N} - \vec X_\text{I} = \frac{m_1 - m_2}{2M_o^2 c^2} \left( \mu v^2 + \frac{\kappa}{r} \right) \vec r
 \end{equation}
It corresponds to the total derivative in rhs of Eq.~(\ref{eq: dR/dt PN eqn 1}).
The other integral of Eqs.~(\ref{eq: dR/dt PN}) and (\ref{eq: dR/dt PN eqn 1}) combined identifies the non-trivial solution
 \begin{equation} \label{eq: R2 PN}
 \vec R_2 = \frac{m_1 - m_2}{2M_o^2 c^2} \mu \left( \vec v \cdot \vec r \right) \vec v
 \end{equation}
The CM-displacement vector $\vec Q$ (\ref{eq: Q R1 R})
 \begin{equation} \label{eq: Q PN}
\vec Q = \vec R_1 - \vec R_2 = \frac{m_1 - m_2}{2 M_o^2 c^2} \left[ \left( \mu v^2 + \frac{\kappa}{r} \right) \vec r - \mu \left( \vec v \cdot
\vec r \right)\vec v \right]
 \end{equation}
is clearly recognized as being proportional to the LRL vector of the
corresponding Newtonian system,
 \begin{equation} \label{eq: Newt RL}
\vec K = \left( \mu v^2 + \frac{\kappa}{r} \right) \vec r - \mu \left( \vec v \cdot
\vec r \right)\vec v = \vec v \times \vec \ell + \frac{\kappa}{r} \vec r \, ,
 \end{equation}
$\vec\ell = \mu \vec r \times \vec v$ being the internal angular momentum vector. This is, in essence, Dahl's result \cite{Dahl97}.

The existence of two independent solutions to Eq.~(\ref{eq: R eqn}) was repeatedly verified, in an analogous manner, for a pair of fully relativistic non-interacting particles \cite{LocCM}; in the post-Newtonian approximation of electromagnetic or gravitational many-body systems \cite{manybodyPN}; and for the post-Newtonian
extensions of general centrally symmetric 2-body systems
\cite{LRLgeneral}. Dahl's results and the cited
computations were all performed in the CM reference frame, but this is only a matter of convenience and simplicity -- it may be explicitly shown that these results are valid in any reference frame, relative to any time-like evolution parameter, as is indeed the case in the two fully relativistic examples discussed in the following. And in all these systems the difference between the two independent solutions, which defines the boost's internal moment, is proportional to a LRL vector with a proportionality
coefficient of the order of $1/c^2$. It follows, therefore, that this is a generic property in relativistic systems.

\section{\label{sec: SVint}CM integration for a two-body system with special scalar-vector interaction}

In the present Section we demonstrate the CM integration for a fully relativistic system with interaction. For simplicity and clarity of the exposition, a relatively simple system, a 2-body system with light-like antisymmetric interaction, is chosen : Events coupled by the interaction satisfy the light-cone condition $\left(x_1- x_2\right)^2 = 0$, so that the  interaction is retarded for one of the particles and advanced for the other. In this way, any event on one particle's trajectory is coupled, via the interaction, to a unique event on the other particle's trajectory, and a canonical structure is possible. Also, it is assumed that the interaction is a combination of vector (EM-like) and scalar interactions whose coupling constants are equal up to a sign. Defining the common coordinates
 \begin{eqnarray} \label{eq: comm coor SV}
 && x^\mu  \equiv x_1^\mu - x_2^\mu \, ,
  \nonumber \\
&& z^\mu \equiv \frac{x_1^\mu + x_2^\mu}{2} + \frac{m_1^2 - m_2^2}{2M^2} \left( x^\mu + 2\frac{P \cdot x}{M^2} P^\mu \right) \, ,
 \end{eqnarray}
such a system was discussed by Duviryak \cite{Duviryak96} who showed that, with the total momentum $P^\mu$ canonically conjugate to $z^\mu$ and $q^\mu$ canonically conjugate to $x^\mu$, the dynamics of the system is determined by the first-class constraint
 \begin{equation} \label{eq: phi SV}
\phi \left( z,P,x,q \right) = q^2 - 2\frac{\left( P \cdot q \right) \left( q \cdot x \right)}{P \cdot x} - \frac{2g\left( M^2 \right)}{P \cdot x} - b\left( M^2 \right) \approx 0
 \end{equation}
Here
 \begin{eqnarray} \label{eq: bM}
b\left(M^2 \right) & \equiv & \frac{M^4 - 2M^2 \left( m_1^2 + m_2^2 \right) + \left( m_1^2 - m_2^2 \right)^2}{4M^2} =
  \nonumber \\
& = & \frac{\left( M^2 - M_o^2 \right) \left[ M^2 - \left( m_1 - m_2 \right)^2 \right]}{ 4M^2}
 \end{eqnarray}
and
 \begin{equation} \label{eq: gM SV}
g\left( M^2 \right) \equiv \frac{\chi \kappa}{2} \left[ M^2 - \left( m_1 - \alpha m_2 \right)^2 \right]
 \end{equation}
where $M_o = m_1 + m_2$, $\chi \equiv {\text{sign}} \left( \dot x_1 \cdot x \right) =
{\text{sign}} \left( \dot x_2 \cdot x \right)$, $\kappa$ the vector
coupling constant and $\kappa ' = \alpha\kappa = \pm \kappa$ the
scalar coupling constant. Assuming a general evolution parameter
$\sigma$, there exists some coefficient $\lambda$ (not necessarily
constant) so that for any observable $A\left( z,P,x,q \right)$ its
$\sigma$-evolution equation is
 \begin{equation} \label{eq: dA/dsigma SV}
\frac{d}{d\sigma} A\left( z,P,x,q \right) = \frac{\lambda}{2} \left\{ A,\phi \right\}
 \end{equation}
with $\left\{ . , . \right\}$ being the canonical Poisson brackets over the 16D phase-space $\left\{ \left( z,P,x,q \right) \right\}$.

The internal momentum of the system is
 \begin{equation} \label{eq: Pi SV}
\Pi^\mu \equiv q^\mu - \frac{P \cdot q}{P \cdot x} x^\mu
 \end{equation}
This is indeed an internal vector, as verified by $\Pi \cdot P = 0$ and $\left\{\Pi^\mu , P^\nu \right\} = 0$. Its evolution equation is
 \begin{equation} \label{eq: dPi/dsigma SV}
\frac{d \Pi ^\mu}{d\sigma} = - \lambda M^2 \frac{g\left( M^2 \right)}{\left( P \cdot x \right)^3} x_\bot^\mu
 \end{equation}
With $\Pi^\mu$ the constraint (\ref{eq: phi SV}) is simplified to
 \begin{equation} \label{eq: phi Pi SV}
\phi = \Pi^2 - \frac{2g\left( M^2 \right)}{P \cdot x} - b\left( M^2 \right) \approx 0
 \end{equation}
$x_\bot^\mu$ and $\Pi^\mu$ form an internal canonical pair, as verified by the equation
 \begin{equation} \label{eq: dxbot/dsigma SV}
\frac{dx_\bot^\mu}{d\sigma} = \frac{\lambda}{2} \frac{\partial \phi}{\partial q^\nu} \Delta^{\mu\nu} = \lambda \Pi^\mu
 \end{equation}
and the fact that $x_\bot^2 = \left( P \cdot x \right)^2 / M^2$. In particular, the internal angular momentum $\ell^{\mu\nu}$ is very conveniently expressed in terms of $x_\bot^\mu$ and $\Pi^\mu$ as
 \begin{equation} \label{eq: ell SV}
\ell^{\mu\nu} = \Delta_\lambda^\mu \Delta_\rho^\nu J^{\lambda\rho} = x_\bot^\mu  \Pi^\nu - x_\bot^\nu \Pi^\mu
 \end{equation}
where
 \begin{equation} \label{eq: J SV}
J^{\mu\nu} = z^\mu P^\nu - z^\nu P^\mu + x^\mu q^\nu - x^\nu q^\mu
 \end{equation}
is the conserved total angular momentum. Also, the centre-of-inertia is found to be
 \begin{equation} \label{eq: XI SV}
X_\text{I}^\mu = - \frac{J^{\mu\nu} P_\nu}{M^2} = z_\bot^\mu + \frac{P \cdot x}{M^2} \Pi^\mu
 \end{equation}

Expressing the Newtonian CM in terms of the canonical variables,
 \begin{eqnarray} \label{eq: Newt CM SV}
 X_N^\mu & = & \frac{m_1 x_1^\mu + m_2 x_2^\mu}{M_o} =
  \nonumber \\
& = & z^\mu + \frac{m_1^2 - m_2^2}{2} \left[ \left( \frac{1}{M_o^2} - \frac{1}{M^2} \right) x^\mu - 2\frac{P \cdot x}{M^4} P^\mu \right] \, ,
 \end{eqnarray}
the CM integration equation (\ref{eq: R eqn}) becomes
 \begin{equation} \label{eq: R eqn1 SV}
\frac{dR^\mu}{d\sigma} = \dot z_\bot^\mu + \frac{m_1^2 - m_2^2}{2} \left( \frac{1}{M_o^2} - \frac{1}{M^2} \right) \dot x_\bot^\mu
 \end{equation}
$z_\bot^\mu$ is not an internal vector. To insure that $R^\mu$ is an internal vector, we use the constancy of $X_\text{I}^\mu$ and obtain from Eq.~(\ref{eq: XI SV})
\[
\dot z_\bot^\mu = - \frac{d}{d\sigma} \left( \frac{P \cdot x}{M^2} \Pi^\mu \right)
\]
so that the equation for $R^\mu$ now becomes
 \begin{equation} \label{eq: R eqn SV}
\frac{dR^\mu}{d\sigma} = \frac{m_1^2 - m_2^2}{2} \left( \frac{1}{M_o^2} - \frac{1}{M^2} \right) \dot x_\bot^\mu - \frac{d}{d\sigma} \left( \frac{P \cdot x}{M^2} \Pi^\mu \right)
 \end{equation}

An immediate solution is, of course,
 \begin{equation} \label{eq: R1 SV}
R^\mu = \frac{m_1^2 - m_2^2}{2} \left( \frac{1}{M_o^2} - \frac{1}{M^2} \right) x_\bot^\mu - \frac{P \cdot x}{M^2} \Pi^\mu
 \end{equation}
which is easily recognized as the trivial solution $R_1^\mu =
X_N^\nu \Delta_\nu^\mu - X_\text{I}^\mu$. To obtain the non-trivial
solution it is convenient to transform Eq.~(\ref{eq: R eqn SV}), using
the constraint equation (\ref{eq: phi Pi SV}), into
 \begin{eqnarray} \label{eq: R eqn2 SV}
\frac{dR^\mu}{d\sigma} && = \frac{m_1^2 - m_2^2}{2M_o^2 M^2} \left( M^2 - M_o^2 \right) \dot x_\bot^\mu - \frac{d}{d\sigma} \left( \frac{P \cdot x}{M^2} \Pi^\mu \right) =
  \nonumber \\
&& = \frac{2\left( m_1 - m_2 \right) b\left( M^2 \right)}{M_o \left[ M^2 - \left( m_1 - m_2 \right)^2 \right]} \dot x_\bot^\mu - \frac{d}{d\sigma} \left( \frac{P \cdot x}{M^2} \Pi^\mu \right) =
  \nonumber \\
&& = \frac{2\left( m_1 - m_2 \right)}{M_o \left[ M^2 - \left( m_1 - m_2 \right)^2 \right]} \left[ \Pi^2 \dot x_\bot^\mu - \frac{2g\left( M^2 \right)}{P \cdot x} \dot x_\bot^\mu  \right] - \frac{d}{d\sigma} \left( \frac{P \cdot x}{M^2} \Pi^\mu \right)
 \end{eqnarray}
With the help of equations (\ref{eq: dPi/dsigma SV}) and (\ref{eq:
dxbot/dsigma SV}) and the relation $\left( P \cdot x \right)^2 =
M^2 x_\bot^2$, the expression within the square brackets is then
converted into a total derivative,
 \begin{eqnarray} \label{eq: PiPix SV}
 \Pi^2 \dot x_\bot^\mu - \frac{2g\left( M^2 \right)}{P \cdot x} \dot x_\bot^\mu & = & \left( \Pi \cdot \dot x_\bot \right) \Pi^\mu - \frac{2g\left( M^2 \right)}{P \cdot x} \dot x_\bot^\mu =
   \nonumber \\
 & = & \frac{d}{d\sigma} \left[ \left( \Pi \cdot x_\bot \right) \Pi^\mu - \frac{g \left( M^2 \right)}{P \cdot x} x_\bot^\mu \right]
 \end{eqnarray}
Thus we are able to identify the second, non-trivial, solution of Eq.~(\ref{eq: R eqn SV}) as
 \begin{equation} \label{eq: R2 SV}
R_2^\mu = \frac{2\left( m_1 - m_2 \right)}{M_o \left[ M^2 - \left( m_1 - m_2 \right)^2 \right]} \left[ \left( \Pi \cdot x_\bot \right) \Pi^\mu - \frac{g\left( M^2 \right)}{P \cdot x} x_\bot^\mu  \right] - \frac{P \cdot x}{M^2} \Pi^\mu
 \end{equation}

To compute the CM-displacement vector $Q^\mu$, it is convenient to transform the trivial solution Eq.~(\ref{eq: R1 SV}), using the constraint equation (\ref{eq: phi Pi SV}), in a way similar to Eq.~(\ref{eq: R eqn2 SV}),
 \begin{eqnarray} \label{eq: R1 2 SV}
R_1^\mu & = & \frac{m_1^2 - m_2^2}{2M_o^2 M^2} \left( M^2 - M_o^2 \right) x_\bot^\mu - \frac{P \cdot x}{M^2} \Pi^\mu =
  \nonumber \\
 & = & \frac{2\left( m_1 - m_2 \right) b \left( M^2 \right)}{M_o \left[ M^2 - \left( m_1 - m_2 \right)^2 \right]} x_\bot^\mu - \frac{P \cdot x}{M^2} \Pi^\mu =
  \nonumber \\
 & = & \frac{2\left( m_1 - m_2 \right)}{M_o \left[ M^2 - \left( m_1 - m_2 \right)^2 \right]} \left[ \Pi^2 x_\bot^\mu - \frac{2g\left( M^2 \right)}{P \cdot x} x_\bot^\mu \right] - \frac{P \cdot x}{M^2} \Pi^\mu
 \end{eqnarray}
Then we obtain
 \begin{eqnarray} \label{eq: Q SV}
Q^\mu = R_1^\mu - R_2^\mu && = \frac{2\left( m_1 - m_2 \right)}{M_o \left[ M^2 - \left( m_1 - m_2 \right)^2 \right]} \left[ \Pi^2 x_\bot^\mu - \left( \Pi \cdot x_\bot \right) \Pi^\mu - \frac{g\left( M^2 \right)}{P \cdot x} x_\bot^\mu \right] =
  \nonumber \\
&& = \frac{2\left( m_1 - m_2 \right)}{M_o \left[ M^2 - \left( m_1 - m_2 \right)^2 \right]} K^\mu
 \end{eqnarray}
where
 \begin{equation} \label{eq: RL SV}
K^\mu \equiv \Pi_\nu \ell^{\mu\nu} - \frac{g\left( M^2 \right)}{P \cdot x} x_\bot^\mu = \Pi^2 x_\bot^\mu - \left( \Pi \cdot x_\bot \right) \Pi^\mu - \frac{g\left( M^2 \right)}{P \cdot x} x_\bot^\mu
 \end{equation}
is the LRL vector of the system \cite{Duviryak96}.

The self Poisson brackets of $K^\mu$ may be computed directly using the canonicity of $x_\bot^\mu$ and $\Pi^\mu$, but this effort may be saved with the help of the LRL symmetry property (\ref{eq: PB KK}), using
 \begin{equation} \label{eq: K^2 SV}
K^2 = b\left( M^2 \right)\ell^2 + \frac{g^2 \left( M^2 \right)}{M^2}
 \end{equation}
Then
 \begin{equation} \label{eq: Pb K SV}
\left\{ K^\mu ,K^\nu \right\} = - \frac{\partial \left( K^2 \right)}{\partial \left( \ell^2 \right)} \ell^{\mu\nu} = - b\left( M^2 \right) \ell^{\mu\nu}
 \end{equation}
The sign of $b\left(M\right)$, which is the same as $\text{sign} \left( M-M_o \right)$, determines the boundness index $\eta$ (Eq.~(\ref{eq: eta})) and thus the type of the symmetry. It is straight-forward to check that the PB in post-Newtonian systems (Eq.(76) of Ref. \onlinecite{LRLgeneral}) is the corresponding limit of Eq.~(\ref{eq: Pb K SV}). A remarkable feature of Eq.~(\ref{eq: Pb K SV}) is its independence on any detail of the interaction, suggesting that it is universal, valid for all 2-body systems.

Finally, we notice that since $x_\bot^\mu$ is the spatial interparticle vector $\vec r$ in the CM frame, the LRL vector (\ref{eq: RL SV}) is of the same structure as the Newtonian LRL vector (\ref{eq: Newt RL}). The similarity implies, in particular, that the orbits are fixed conic sections. This simplicity is due to the equality (up to sign) of the coupling constants of the scalar and vector interactions. To elucidate this aspect, a simplified version of this system -- a particle in a Coulomb field modified by a scalar field -- is discussed in Appendix \ref{sec: appa}.

\section{\label{sec: non-int many}Centre-of-mass integration for non-interacting fully relativistic many-body system}

A main argument that follows from the discussion so far in the present work is that LRL vectors are derived from the Lorentz boost. Although the LRL symmetry is known so far to be found only in 2-body systems, the fact that any composite system is endowed with a Lorentz boost strongly suggests that such systems could also be endowed with LRL symmetry. This possibility has already been explored in the post-Newtonian approximation\cite{manybodyPN}. In the present Section we compute the internal moment of the Lorentz boost (or, what's equivalent, the shift vector $Q^\mu$) for a fully relativistic, non-interacting many-body system and demonstrate that it is indeed a LRL vector. This may serve as a starting point for establishing the relativistic LRL symmetry in general many-body systems in the future.

Consider a system of free, non-interacting $N$ point particles
with masses $m_a$, moving on straight-line trajectories
$x_a^\mu(\tau_a)$ with constant unit velocities $u_a^\mu$ and
$\tau_a$ being the proper time of the $a$-th particle. The total
linear and angular momenta are
 \begin{equation} \label{eq: P & J free}
P^\mu = \sum\limits_a p_a^\mu \hskip 50pt J^{\mu\nu} =
\sum\limits_a \left( x_a^\mu p_a^\nu - x_a^\nu p_a^\mu \right)
 \end{equation}
with the single particles' linear momenta $p_a^\mu = m_a u_a^\mu$. A generalized
Lorentz factor is defined by
 \begin{equation} \label{eq: gamma_a def}
 \gamma_a \equiv \left( \frac{d\tau_a}{d\sigma} \right)^{-1} = \frac{1}{\sqrt{-{\dot{x}_a}^2}} \, ,
 \end{equation}
allowing us to write the particles' generalized velocities as
 \begin{equation} \label{eq: gen vel}
 \dot{x}^\mu = \gamma_a^{-1} u_a^\mu
 \end{equation}

Let $E_a = - p_a \cdot U$ be the single particle energy
in the CM-frame. It is convenient to introduce the notations
 \begin{equation} \label{eq: def xia}
\xi_a^\mu \equiv \Delta_\nu^\mu x_a^\nu = x_a^\mu + \frac{(x_a \cdot P)}{M^2} P^\mu
 \end{equation}
for the single particle spatial coordinate in the CM reference
frame, and
 \begin{equation} \label{eq: def qa}
q_a^\mu \equiv \Delta_\nu^\mu p_a^\nu = p_a^\mu - E_a U^\mu
 \end{equation}
for the spatial component, in the CM reference frame, of the
single particle momentum, satisfying the CM-constraint
 \begin{equation} \label{eq: sum qa}
  \sum\limits_a q_a^\mu = 0
 \end{equation}
The single particle energies in the CM frame then become $E_a = \sqrt {m_a^2 + q_a^2}$ and the internal angular momentum is $\ell^{\mu\nu} = \sum_a \left( \xi_a^\mu q_a^\nu - \xi_a^\nu q_a^\mu \right)$.

Substituting $P^\mu$ and $J^{\mu\nu}$ from Eq.~(\ref{eq: P & J free}),
the centre-of-inertia (\ref{eq: XI def}) is
 \begin{equation} \label{eq: RI non-int}
X_\text{I}^\mu = - \frac{J^{\mu\nu} P_\nu}{M^2} =
\sum\limits_a \left[ \frac{E_a}{M} x_a^\mu + \frac{\left(x_a \cdot P
\right)}{M^2} p_a^\mu \right] = \sum\limits_a \left[ \frac{E_a}{M} \xi_a^\mu + \frac{\left(x_a \cdot P \right)}{M^2} q_a^\mu
\right]
 \end{equation}
with the trivial solution to Eq.~(\ref{eq: R eqn})
 \begin{equation} \label{eq: R1 free}
R_1^\mu = \frac{\sum\nolimits_a m_a \xi_a^\mu}{M_o}
- X_\text{I}^\mu = \sum\limits_a \left( \frac{m_a}{M_o} - \frac{E_a}{M}
\right) \xi_a^\mu - \sum\limits_a \frac{\left(x_a \cdot P \right)}{M^2} q_a^\mu
 \end{equation}

With the relation (\ref{eq: gen vel}), the fundamental equation of the
relativistic CM integration, Eq.~(\ref{eq: R eqn}), becomes
 \begin{equation} \label{eq: R eqn2}
 \frac{d R^\mu}{d\sigma} = \frac{\sum_a m_a \dot{\xi}^\mu_a}{M_o} =
\frac{\sum_a \gamma^{-1}_a q_a^\mu}{M_o}
 \end{equation}
Using the fact that the lorentz factor also satisfies $\gamma_a^{-1} = -u_a
\cdot \dot{x}_a$, and since the particles' unit velocities $\{u_a^\mu\}$
and all the $q_a^\mu$ are constant in the absence of interactions, it can be shown (see Appendix \ref{sec: appb} for details) that a non-trivial solution is possible only if there exists a vector $G^\mu$, composed of the particles' unit velocities, that satisfies
 \begin{equation} \label{eq: gamma to G}
\sum\limits_a \gamma_a^{-1} q_a^\mu = \sum\limits_a \left( G \cdot
\dot x_a \right) q_a^\mu
 \end{equation}
so that
 \begin{equation} \label{eq: R2 G}
R_2^\mu = \frac{1}{M_o} \sum\limits_a \left( G \cdot x_a
\right)q_a^\mu
 \end{equation}
Since Eq.~(\ref{eq: gamma to G}) is linear in the particles' velocities
$\dot x_a^\mu$, and these velocities are functionally independent
because the particles' trajectories are independent, it is
satisfied, again applying Eq.~(\ref{eq: gen vel}), only if
 \begin{equation} \label{eq: G ua}
 G \cdot u_a = 1 \qquad \forall a
 \end{equation}
Writing $G^\mu$ as the linear combination $G^\mu = \sum_a \alpha_a
u_a^\mu$ with constant coefficients $\alpha_a$, these coefficients
are uniquely determined by the linear system
$$
\sum\limits_b \left( u_a \cdot u_b \right)\alpha_b = 1 \qquad
\forall a
$$
thus verifying that $G^\mu$ is uniquely determined by Eq.~(\ref{eq: G
ua}).

The non-trivial solution (\ref{eq: R2 G}) contains both spatial and
time-like parts of the particles' coordinates $x_a^\mu$ in one
term. To separate them, in analogy with $R_1^\mu$ in Eq.~(\ref{eq: R1 free}), we notice first that multiplying the $a$-th equation in
(\ref{eq: G ua}) by $m_a$ and summing them all yields
 \begin{equation} \label{eq: G P}
 G \cdot P = M_o
 \end{equation}
Separating from $G^\mu$ its CM-spatial part,
 \begin{equation} \label{eq: G sep P}
 G^\mu = G_\bot^\mu - \frac{G \cdot P}{M^2} P^\mu = G_\bot^\mu - \frac{M_o}{M^2} P^\mu \, ,
 \end{equation}
the non-trivial solution (\ref{eq: R2 G}) becomes
 \begin{equation} \label{eq: R2 free}
R_2^\mu = \frac{1}{M_o} \sum\limits_a \left( G_\bot \cdot \xi_a
\right) q_a^\mu - \sum\limits_a \frac{x_a \cdot P}{M^2} q_a^\mu
 \end{equation}

The trivial solution $R_1^\mu$ (\ref{eq: R1 free}) may also be expressed in terms of the vector $G^\mu$. The coefficients of $\xi_a^\mu$ may be expressed in terms of $G_\bot^\mu$ via the relations
 \begin{equation} \label{eq: G dot q}
\frac{G_\bot \cdot q_a}{M_o} = \frac{G_\bot \cdot p_a}{M_o} = \frac{G \cdot p_a}{M_o} +
\frac{P \cdot p_a}{M^2} = \frac{m_a}{M_o} - \frac{E_a}{M}
 \end{equation}
so that $R_1^\mu$ becomes
 \begin{equation} \label{eq: R1 non-int G}
R_1^\mu = \frac{1}{M_o} \sum\limits_a \left( G_\bot \cdot q_a \right) \xi_a^\mu - \sum\limits_a \frac{x_a \cdot P}{M^2} q_a^\mu
 \end{equation}
Combining Eqs.~(\ref{eq: R1 non-int G}) and (\ref{eq: R2 free})
for the computation of the shift vector $Q^\mu$, the terms
containing the time-like part of the particles' coordinates cancel
each other and we obtain
 \begin{eqnarray} \label{eq: Q free}
 Q^\mu & = & R_1^\mu - R_2^\mu = \frac{1}{M_o} \sum\limits_a \left[ \left( G_\bot \cdot q_a \right) \xi_a^\mu - \left( G_\bot \cdot \xi_a \right) q_a^\mu \right] =
\nonumber \\
& = & \frac{1}{M_o} G_{\bot \nu} \sum\limits_a \left( q_a^\nu \xi_a^\mu - \xi_a^\nu  q_a^\mu \right) = \frac{G_\nu \ell^{\mu \nu}}{M_o}
 \end{eqnarray}

$Q^\mu$ is indeed a generalized LRL vector. For a 2-body
system let $q^\mu \equiv q_1^\mu = -q_2^\mu$. Then $G_\bot^\mu$ is
easily found as
\begin{equation} \label{eq: G 2}
G_\bot^\mu = \frac{2\left( m_1 - m_2 \right)}{M^2 - \left( m_1 - m_2 \right)^2} q^\mu
\end{equation}
so that $Q^\mu$ is proportional to $q_\nu \ell^{\mu \nu}$, a more
familiar form of the interaction-free 2-body LRL vector.
For a many-body system the vector (\ref{eq: Q free}) satisfies
$Q^\mu \ell_\mu = 0$, and its self PB are found, taking into account
that $\left\{G^\mu , G^\nu\right\} = 0$, by direct computation,
\begin{equation} \label{eq: PB QQ free}
 \left\{Q^\mu , Q^\nu\right\} = \frac{1}{M_o^2} \left\{G_\lambda \ell^{\mu \lambda} , G_\rho \ell^{\nu \rho}\right\}
 = - \frac{G_\bot^2}{M_o^2} \ell^{\mu\nu}
\end{equation}
On the other hand, using the identity $\ell^{\mu \lambda}\ell_{\nu \lambda} = \ell^2 \Delta^\mu_\nu - \ell^\mu \ell_\nu$ it follows that $Q^2 = \left[G_\bot^2 \ell^2 - \left( G \cdot \ell\right)^2\right]/M_o^2$. From $\left\{G^\mu , G^\nu\right\} = 0$ it follows that $\left\{Q^\mu , G \cdot \ell \right\} = 0$. Thus with ${\mathcal A} = G \cdot \ell$, $Q^\mu$ satisfies the conditions of proposition 2 and the self PB (\ref{eq: PB QQ free}) are verified by Eq.~(\ref{eq: PB KK}). This completes the proof that $Q^\mu$ is indeed (proportional to) a LRL vector, and the many-body system enjoys LRL symmetry.

\section{\label{sec: CMloc}The shift vector and CM localizability}

With $j^{\mu\nu}$ being decomposed as in Eq.~(\ref{eq: j decomposion}), the vector $Q^\mu$ may also be regarded as generating the internal Lorentz boost relative to the CM reference frame. The possibility of splitting $J^{\mu\nu}$, as in Eq.~(\ref{eq: split J}), was appreciated, at least in principle, from the early days of the search for the relativistic CM \cite{LoRom74}, with various propositions for what should be, in the notation of the present article, the vector $Q^\mu$ \cite{Pryce,Fleming65a}. So far in the literature an internal moment is associated with the Lorentz boost only if internal spin is assumed to exist as an independent entity, in which case the moment depends on the spin. Here, for the first time, the association of the internal Lorentz boost with the LRL vector was demonstrated and established.

The preceding analysis distinguished the vectors $R_1^\mu$ and $R_2^\mu$ as solutions of Eq.~(\ref{eq: R eqn}). Although, as integrals, the solutions of Eq.~(\ref{eq: R eqn}) are defined up to an arbitrary additive constant, these particular solutions have characteristics that distinguish them from arbitrary integrals : Eq.~(\ref{eq: dR/dt PN eqn 1}) shows explicitly that Eq.~(\ref{eq: R eqn}) leads to two specific solutions, one which is proportional to the relative coordinates and another which is proportional to the relative velocities or momenta. The first one was identified as the vector $R_1^\mu$ defined in Eq.~(\ref{eq: R1 def}), while the other was referred to as the non-trivial solution $R_2^\mu$. This characteristic behaviour may be verified for $R_1$ in equations (\ref{eq: R1 PN}), (\ref{eq: R1 SV}) and (\ref{eq: R1 free}), with the leading term being proportional to $\vec r$, $x_\bot^\mu$ and $\xi_a^\mu$ respectively, and for $R_2$ in equations (\ref{eq: R2 PN}), (\ref{eq: R2 SV}), and (\ref{eq: R2 free}), with the leading term being proportional to $\vec v$, $\Pi^\mu$ and $q_a^\mu$ respectively. This is also the case in systems discussed elsewhere \cite{LocCM,LRLgeneral,manybodyPN}.

These two characteristic solutions define the characteristic shift vector $Q^\mu = R_1^\mu - R_2^\mu$, which in turn defines the internal moment of the Lorentz boost relative to the CM frame. A remarkable feature of this $Q^\mu$ is its being exactly proportional to the LRL vector of the system, thus pointing (in 2-body systems) in the direction of closest approach (generalized perihelion). This has been verified for various systems, both in the present paper (Eqs.~(\ref{eq: Q PN}) and (\ref{eq: Q SV})) and elsewhere \cite{LRLgeneral}. The proportionality of $Q^\mu$ to the LRL vector $K^\mu$ is a very interesting aspect of the Lorentz boost, because $Q^\mu$ could be $Q^\mu = \alpha(M,\ell^2) K^\mu + \beta(M,\ell^2) {\ell^\mu}_\nu K^\nu$ with arbitrary coefficients $\alpha(M,\ell^2)$ and $\beta(M,\ell^2)$ and still maintain the same internal symmetry. The reason for this particular proportionality is not clear yet.

The existence of more than one independent solution to Eq.~(\ref{eq: R eqn}) also adds new insight into the long-standing issue of the localizability of the relativistic CM coordinate. With the two special independent solutions $R_1^\mu$ and $R_2^\mu$ the CM coordinate may be either the inertia centroid (Eqs.~(\ref{eq: X trajectory}) and (\ref{eq: XI def}), combined)
 \begin{equation} \label{eq: inertia centro}
 X_1^\mu(\tau) = - \frac{J^{\mu\nu} P_\nu}{M^2} + \frac{P^\mu}{M} \cdot \tau
 \end{equation}
or removed from it by the shift vector $Q^\mu$,
 \begin{equation} \label{eq: shift centro}
 X_2^\mu(\tau) = - \frac{J^{\mu\nu} P_\nu}{M^2} + Q^\mu + \frac{P^\mu}{M} \cdot \tau
 \end{equation}
Consequently, the CM coordinate cannot, in principle, be uniquely defined in a point-like manner, with $Q^\mu$ providing a measure of its non-uniqueness.

$Q^\mu$ is directed towards a point of closest approach. For unbound systems (collisions) there is just one such point, so that $Q^\mu$, in this sense, is unique. For general bound systems there is an infinite number of directions of closest approach, with corresponding infinite possible directions for $Q^\mu$. Thus we should consider a \textit{'centre-of-mass' domain} relative to the centre-of-inertia, which is linear in the case of unique direction of closest approach, or circular in the case of multiple directions of closest approach.

This non-uniqueness of the relativistic CM and its dependence on $Q^\mu$ may be understood in the following way \cite{LocCM}. The Newtonian CM coordinate $\vec X_\text{N}$ depends only on the particles' coordinates and masses. Thus, any value may be attached to it regardless the actual configuration in which the system is, and this value may remain unchanged even when the configuration (namely, internal energy and/or angular momentum) changes. On the other hand, the relativistic centre-of-inertia depends on the system's configuration via the particles' energies, and is therefore sensitive to any changes in the configuration which are generated by the boost's internal moment. The uniqueness of the Newtonian CM may therefore be regarded as reflecting its independence on the system's configuration, with the opposite case for the relativistic CM. The only exception is the case of equal masses, in which the centre-of-mass must be midway between the two particles regardless of their configuration, and indeed, as is evident from Eqs.~(\ref{eq: Q PN}), (\ref{eq: Q SV}) and (\ref{eq: G 2}), $Q^\mu$ vanishes in this case.

In Lie-Poisson algebraic terms, in analogy with quantum mechanics, localizability is synonym with requiring the relativistic CM coordinate to be canonical, with vanishing self PB. Assuming the fundamental PB $\left\{X^\mu,P^\nu \right\} = g^{\mu\nu}$, it is not difficult to show that the self PB of the CM coordinates satisfy
 \begin{equation} \label{eq: Del X-X PB}
\Delta_\lambda^\mu \Delta_\rho^\nu \left\{ X^\lambda,X^\rho \right\} = \left\{ Q^\mu,Q^\nu \right\} + \frac{\ell^{\mu\nu}}{M^2}
 \end{equation}
The vanishing of the rhs, implying the self PB of the shift vector $Q^\mu$ being
 \begin{equation} \label{eq: Q-Q PB}
\left\{ Q^\mu , Q^\nu \right\} = -\frac{\ell^{\mu\nu}}{M^2}
 \end{equation}
is therefore a necessary condition for $X^\mu$ being localizable. However, combining Eqs.~(\ref{eq: Q SV}) and (\ref{eq: Pb K SV}), the self PB of $Q^\mu$ there is
\begin{equation} \label{eq: PB Q-Q rel}
\left\{ Q^\mu , Q^\nu \right\} = -\frac{4\left( m_1 - m_2 \right)^2 b\left(M\right) }{M_o^2 \left[M^2 - \left( m_1 - m_2 \right)^2\right]^2} \ell^{\mu\nu}
\end{equation}
and it is easily verified that the same result (in the appropriate limit) is also obtained for post-Newtonian systems. Moreover, the PB (\ref{eq: Q-Q PB}) imply the boundness index (Eq.~(\ref{eq: eta})) $\eta = -1$ which consequently implies that the system is necessarily unbound. Thus, even if we tried to scale $Q^\mu$ to $Q'^\mu = \alpha \left(R_1^\mu - R_2^\mu\right)$ with $\alpha$ some appropriately chosen coefficient so that $Q'^\mu$ satisfies Eq.~(\ref{eq: Q-Q PB}), that could be possible only for unbound systems since $b\left(M^2\right) \propto M - M_o$. In the general case, therefore, $X^\mu$ is not expected to be canonical, thus localizable.

This situation is somewhat similar to the one encountered with spinning point particles, even at the classical (non-quantum) level. It is well-known \cite{Corben68,KudryObukhov10} that such particles, endowed with conserved linear momentum $p^\mu$ and angular momentum $J^{\mu\nu}$, and identifying the relativistic spin tensor as $s^{\mu\nu} = J^{\mu\nu} - x^\mu p^\nu + x^\nu p^\mu$ and the bare mass as $m = \sqrt{-p_\mu p^\mu}$, don't move on the inertia centroid (as in Eq.~(\ref{eq: inertia centro})) but rather rotate around it, thus exhibiting a helical motion in Minkowski space-time. The rotation is encapsulated in the uniformly rotating shift vector $q^\mu$, defined in complete analogy with Eq.~(\ref{eq: Q from j}) as $q^\mu \equiv s^{\mu\nu} p_\nu / m^2$. The difference between the two situations is that for spinning point particles the spin condition $s^{\mu\nu} \dot x_\nu = 0$ is assumed, with the effect of allowing  only helical orbits and limiting the number of independent spin components to three only, while in the preceding analysis of composite systems there is no such condition on the internal angular momentum $j^{\mu\nu}$ (allowed to have 6 independent components), and the CM coordinate moves on the centroid (\ref{eq: shift centro}), shifted from the inertia centroid by the constant vector $Q^\mu$.

\section{\label{sec: concrem}Concluding remarks}

We have demonstrated, as the main subject of the present article, that Lorentz-Poincar\'{e} symmetry implies not only internal rotational symmetry but also the existence of LRL symmetry in composite relativistic systems. Together, the internal rotational symmetry and the LRL symmetry form the fundamental internal symmetry of all relativistic systems.

The association of the LRL vector with the internal Lorentz boost finds its significance in the symmetries that these objects generate : The (global) Lorentz boost changes the state of motion -- the way particles move -- relative to the (external) reference frame. In a similar way, the internal boost changes the state of motion -- the internal configuration -- relative to the centre-of-mass. Spatial rotations and the Lorentz boost form together the generalized rotations in Minkowski space-time. Thus, the internal boost is related to the global Lorentz boost in much the same way as internal rotations are related to the global rotations. It is no surprise, therefore, that the boost's internal moment is identified with the LRL symmetry, because classically it is the LRL vector that generates the
transformations that change the internal configuration of the system \cite{LRLgeneral}.

To put it differently, the LRL symmetry is the internal aspect generically associated with the (global) Lorentz transformations, in the same way that the internal, spatial rotations are the internal aspect of the global rotations (the latter being defined relative to a fixed frame of reference); and the rotational and LRL symmetries are attached together internally in the same way that global rotations and Lorentz transformations form together the generalized rotations in Minkowski space-time. These relations are illustrated in the following diagram \cite{LRLgeneral} :
\[
\begin{array}{*{20}{c}}
 {} & \vline  & \begin{array}{c}
\textbf{rotations}
\end{array}
 & + &
 \begin{array}{c}
\textbf{change of} \\ \textbf{configuration}
\end{array}
 \\
\hline
 \textbf{Global space-time symmetry :} \, & \vline &
\begin{array}{c}
\textrm{global} \\ \textrm{rotations}
\end{array}
 & \, + \, &
 \begin{array}{c}
\textrm{Lorentz} \\ \textrm{transformations}
\end{array}
 \\
 {} & \vline  & \updownarrow &  & \updownarrow  \\
 \textbf{Internal symmetry :} \, & \vline &
\begin{array}{c}
\textrm{internal} \\ \textrm{rotations}
\end{array}
 &  \, + \, &
 \textrm{LRL}
\end{array}
\]
The LRL symmetry is therefore geometrical in nature, as the LRL vector is associated with the internal moment (relative to the centre-of-mass) corresponding to the Lorentz boost. LRL vectors and LRL symmetry are therefore universal, characteristic of all relativistic systems.

In conclusion, Lorentz-Poincar\'{e} symmetry demands the extended internal symmetry, manifested via the corresponding internal moments and closely associated with the determination of the relativistic centre-of-mass, providing us with new tools for the investigation of the internal dynamics of relativistic systems. As a consequence, the CM coordinate of composite relativistic systems should be explicitly constructed not only of the global quantities $P^\mu$ and $J^{\mu\nu}$, but
also of the LRL internal dynamics of the system.

\appendix

\section{\label{sec: appa}The relativistic scalar-Coulomb case with infinite central mass}

It was noted at the end of section~\ref{sec: SVint} that the LRL vector (\ref{eq: RL SV}) is of the same structure as the Newtonian LRL vector (\ref{eq: Newt RL}). This simplicity is due to the equality (up to sign) of the coupling constants of the scalar and vector interactions. To elucidate this aspect, let us consider a relativistic particle with variable mass
 \begin{equation} \label{eq: vari m}
\mu^2(r) = m^2 + \frac{\kappa'^2}{r^2}
 \end{equation}
in a Coulomb field with fixed centre (infinite central mass), with the Hamiltonian
 \begin{equation} \label{eq: H Coul}
 H = \sqrt {p^2 + m^2 + \frac{\kappa'^2}{r^2}} + \frac{\kappa}{r}
 \end{equation}
in the CM frame. In a configuration with given energy $H = E$ and internal angular momentum $\vec \ell = \vec r \times \vec p$, we isolate the squared momentum as
 \begin{equation} \label{eq: p2 SV}
p^2 = p_r ^2 + \frac{\ell^2}{r^2} = \left( E - \frac{\kappa}{r} \right)^2 - m^2 - \frac{\kappa'^2}{r^2} = E^2 - m^2 - \frac{2\kappa E}{r} - \frac{\kappa'^2 - \kappa^2}{r^2}
 \end{equation}
Equating now, up to a sign, the coupling constants, $\kappa = \pm \kappa'$, results in the canceling of the $1/r^2$ terms on the rhs of Eq.~(\ref{eq: p2 SV}), keeping the centrifugal term $\ell^2 / r^2$ as in Newtonian mechanics and leading to the equation
 \begin{equation} \label{eq: sC eqn}
\left( \frac{1}{r^2} \frac{dr}{d\theta} \right)^2 + \left( \frac{1}{r} + \frac{\kappa E}{\ell^2} \right)^2 = \frac{\left( E^2 - m^2 \right)\ell^2 + E^2 \kappa^2}{\ell^4}
 \end{equation}
with the solution
 \begin{equation} \label{eq: orbit Coul}
 \frac{1}{r} + \frac{\kappa E}{\ell^2} = \frac{\sqrt{\left( E^2 -
m^2 \right)\ell^2 + E^2 \kappa^2}}{\ell^2} \cos \theta
 \end{equation}
The solution is a fixed, non-rotating, conical curve, as for the non-relativistic Coulomb case.

The similarity in form of the solutions implies similarity of the LRL vectors, which is in the present case
 \begin{equation} \label{eq: vec K sC}
 \vec K \equiv \vec \ell \times \vec p - \frac{\kappa E}{r} \vec r
 \end{equation}
It is a constant vector with magnitude
 \begin{equation} \label{eq: K mag sC}
 \left| \vec K \right|^2 = \left(E^2 - m^2 \right) \ell^2 + E^2 \kappa^2 \, ,
 \end{equation}
directed along the major axis of the conic section. It would be interesting to study the quantization of this vector, because, unlike the non-relativistic case, here it depends on energy eigenvalues, both positive and negative.

Evidently, if $\kappa \ne \pm \kappa'$ then the extra $1/r^2$ term on the rhs of Eq.~(\ref{eq: p2 SV}) would modify the centrifugal term and lead to a rotating conic section with the corresponding complicated LRL vector, as in the relativistic Kepler-Coulomb system with fixed centre \cite{Yoshida88b,LLFields75,LRLgeneral}, or the Newtonian Kepler-Coulomb system modified with $1/r^2$ potential \cite{Heintz76}.

\section{\label{sec: appb}Proof of the uniqueness of the solution Eq.~(\ref{eq: R2 G})}

In the following we prove that Eq.~(\ref{eq: gamma to G}) is necessary
for an internal non-trivial solution $R_2^\mu$. In an $N$-body
system, let $1 \le m \le N$. Then, by means of the CM-constraint
(\ref{eq: sum qa}), $q_m^\mu$ is eliminated from the sum $\sum_a
\gamma_a^{-1} q_a^\mu$ in Eq.~(\ref{eq: R eqn2}) to yield
 \begin{equation} \label{eq: gamma N-1 q}
\sum\limits_a \gamma_a^{-1} q_a^\mu = \sum\limits_{a \ne m} \left(
\gamma_a^{-1} - \gamma_m^{-1} \right)q_a^\mu
 \end{equation}
Since all the $N-1$ remaining $q_a^\mu$'s are dynamically
independent, and since the sum $\sum_a \gamma_a^{-1} q_a^\mu$ must
lead to an internal integral, then, taking into account the
identity  $\gamma_a^{-1} = -u_a \cdot \dot{x}_a$, it follows that
for each of the coefficients in Eq.~(\ref{eq: gamma N-1 q}) there must
be a vector $G_a^{\left( m \right)\mu}$, independent of the
particles' coordinates and depending only on the particles' unit
velocities $u_a^\mu$, so that $\gamma_a^{-1} - \gamma_m^{-1} =
G_a^{\left( m \right)} \cdot \left( \dot{x}_a - \dot{x}_m
\right)$.

The use of the particular index $m$ in the resultant sum
 \begin{equation} \label{eq: sum no qm}
\sum\limits_a \gamma_a^{-1} q_a^\mu = \sum\limits_{a \ne m}
G_a^{\left( m \right)} \cdot \left( \dot{x}_a - \dot{x}_m \right)
q_a^\mu
 \end{equation}
is of course arbitrary, and another index $1 \le n \le N,n \ne m$
could be used instead with the sum
 \begin{equation} \label{eq: sum no qn}
\sum\limits_a \gamma_a^{-1} q_a^\mu = \sum\limits_{a \ne n}
G_a^{\left( n \right)} \cdot \left( \dot{x}_a - \dot{x}_m \right)
q_a^\mu
 \end{equation}
Eliminating $q_m^\mu$ from the sum in Eq.~(\ref{eq: sum no qn}), the
latter becomes after some algebra
 \begin{eqnarray} \label{eq: sum m n}
 \sum\limits_a \gamma_a^{-1} q_a^\mu && = \sum\limits_{a \ne n,m} {\left[ {G_a^{\left( n \right)}  \cdot \dot{x}_a  - G_m^{\left( n \right)}  \cdot \dot{x}_m  + \left( {G_m^{\left( n \right)}  - G_a^{\left( n \right)} } \right) \cdot \dot{x}_n } \right]q_a^\mu  } +
\nonumber \\
 && \hskip60pt + G_m^{\left( n \right)}  \cdot \left( {\dot{x}_n  - \dot{x}_m } \right)q_n^\mu
 \end{eqnarray}

The sums in Eqs.~(\ref{eq: sum no qm}) and  (\ref{eq: sum m
n}), must be identical, but their identity can be realized iff all
the vectors $G_a^{\left( m \right)\mu}$ are identical,
$G_a^{\left( m \right)\mu} = G^\mu  \quad  \forall m,a$. This
completes the proof of the uniqueness of the form (\ref{eq: R2 G})
for the non-trivial solution.

\vskip30pt

\rule{10cm}{1pt}

%%%%%%%%%%%%%%%%%%%%%%%%%%%%%%%%%%%%%%%%%%%%%%%%%%%%%%%%%%%%%%%%%%%%%%%%%
%References:
%%%%%%%%%%%%%%%%%%%%%%%%%%%%%%%%%%%%%%%%%%%%%%%%%%%%%%%%%%%%%%%%%%%%%%%%%

    \end{document}